\documentclass[aps,prd,10pt,notitlepage,superscriptaddress,nofootinbib,numbers]{revtex4-2}

\usepackage[utf8]{inputenc}
\usepackage[T1]{fontenc}
\usepackage{lmodern}
\usepackage{microtype}
\usepackage{amsmath,amssymb,amsfonts,mathrsfs}
\usepackage{bm}
\usepackage{graphicx}
\usepackage{float}
\usepackage[dvipsnames]{xcolor}
\usepackage{hyperref}
\hypersetup{colorlinks=true,citecolor=Purple,linkcolor=Purple,urlcolor=Purple,breaklinks=true}
\usepackage[capitalize]{cleveref}
\usepackage{tikz}

\definecolor{lime}{HTML}{A6CE39}
\DeclareRobustCommand{\orcidicon}{%
	\begin{tikzpicture}
		\draw[lime, fill=lime] (0,0) 
		circle [radius=0.16] 
		node[white] {{\fontfamily{qag}\selectfont \tiny ID}};
		\draw[white, fill=white] (-0.0625,0.095) 
		circle [radius=0.007];
	\end{tikzpicture}
	\hspace{-2mm}
}

\foreach \x in {A, ..., Z}{%
	\expandafter\xdef\csname orcid\x\endcsname{\noexpand\href{https://orcid.org/\csname orcidauthor\x\endcsname}{\noexpand\orcidicon}}
}

\newcommand\orcidJonathan{{\href{https://orcid.org/0000-0001-9291-0893}{\orcidicon}}}
\newcommand\orcidcelio{{\href{https://orcid.org/0000-0002-1266-2218}{\orcidicon}}}
\newcommand\orcidmessias{{\href{https://orcid.org/0000-0002-6270-6207}{\orcidicon}}}
\newcommand\orcidraissa{{\href{https://orcid.org/0000-0002-9353-0101}{\orcidicon}}}

\begin{document}

\title{Gravitational Lensing in a Kasner Background: Distinguishing Wormholes and Black Holes}

\author{Celio R. Muniz\orcidcelio}
\email{celio.muniz@uece.br}
\affiliation{Universidade Estadual do Cear\'a (UECE), Faculdade de Educa\c{c}\~ao, Ci\^encias e Letras de Iguatu, Av. D\'ario Rabelo s/n, Iguatu - CE, 63.500-00 - Brasil}
\author{Jonathan A. Rebouças\orcidJonathan}
\email{jalvesreboucas@ifce.edu.br}
\affiliation{Instituto Federal de Educação Ciências e Tecnologia do Ceará (IFCE), Iguatu-CE, Brazil}
\author{M. B. Cruz\orcidmessias}
\email{messiasdebritocruz@servidor.uepb.edu.br}
\affiliation{Universidade Estadual da Para\'iba (UEPB), \\ Centro de Ci\^encias Exatas e Sociais Aplicadas (CCEA), \\ R. Alfredo Lustosa Cabral, s/n, Salgadinho, Patos - PB, 58706-550 - Brazil.}
\author{R. M. P. Neves\orcidraissa}
\email{raissa.pimentel@uece.br}
\affiliation{Universidade Estadual do Cear\'a (UECE), Faculdade de Educa\c{c}\~ao, Ci\^encias e Letras de Iguatu, Av. D\'ario Rabelo s/n, Iguatu - CE, 63.500-00 - Brazil.}

\begin{abstract}

We investigate gravitational lensing by compact objects embedded in anisotropic Bianchi-I cosmologies using directional Jacobi maps within the thin-lens approximation. The formalism is developed for a general diagonal Bianchi-I spacetime and specialized to the Kasner solution as an analytically tractable background. Using the Ellis--Bronnikov wormhole and the Schwarzschild black hole as representative lenses, we derive anisotropic lens equations, characteristic axis-aligned lensing scales, and the corresponding critical curves. We show that the directional splitting of the characteristic scales depends on the complete source--lens--observer optical propagation and provides a geometric probe of anisotropic expansion independent of the overall lens scale. By contrast, the exact critical curves exhibit a much weaker deformation, indicating that characteristic-scale splitting and critical-curve morphology probe distinct aspects of the lens mapping. The comparison between wormhole and black-hole lenses further reveals that identical anisotropic backgrounds are filtered differently by distinct weak-field deflection laws. These results provide a simple framework for disentangling cosmological anisotropy from the local geometry of compact lenses.

\end{abstract}

\maketitle

\tableofcontents

\section{Introduction}\label{sec:introduction}

Gravitational lensing translates the propagation of null geodesics into directly interpretable observables, including multiple images, Einstein rings, magnification patterns, critical curves, caustics, and weak distortions of extended sources. Since the earliest point-mass treatment of aligned lensing, the subject has developed into a central tool for studying compact objects, testing relativistic gravity, and mapping the matter distribution and geometry of the Universe \cite{Einstein1936LensAction,Schneider:1992bge,Wambsganss1998Astronomy,Bartelmann2001Weak,Perlick:2004tq,Petters2010Mathematics}. A lensing configuration is nevertheless not determined by the local deflector alone. The observed mapping also depends on the optical propagation from the observer to the lens and from the lens to the source. Standard applications encode this propagation through scalar angular-diameter distances computed in an isotropic Friedmann--Lema\^itre--Robertson--Walker (FLRW) background. This reduction is exceptionally successful, but it can conceal directional information whenever the cosmological geometry itself is anisotropic.

Cosmological isotropy is therefore both a simplifying principle and an empirical statement. Spatially homogeneous Bianchi models provide the canonical setting in which departures from isotropic expansion can be formulated without abandoning large-scale homogeneity \cite{Ellis1969Homogeneous,Jacobs1968BianchiI,Wainwright1997Dynamical}. Within this class, Bianchi-I spacetime is the spatially flat case with three independent directional scale factors, while the vacuum Kasner solution supplies its simplest exact power-law realization \cite{Kasner:1921gyt}. Kasner epochs also arise as elementary pieces of the more general anisotropic dynamics near spacelike singularities \cite{Belinskii1970Oscillatory}, whereas questions of isotropization and the dynamical status of nearly Friedmann universes have long motivated the systematic study of homogeneous shear \cite{Collins1973Isotropic}. These results make Kasner geometry a natural analytic laboratory: it is not a realistic model of the late-time Universe, but it isolates expansion and contraction along inequivalent axes in a form that remains exactly tractable.

The observational relevance of anisotropy lies in the fact that it modifies redshift, apparent size, luminosity distance, and image distortion in a direction-dependent way. Covariant observational relations in general spacetimes were developed early in relativistic cosmology \cite{Kristian1966Observations}, and explicit studies of shearing homogeneous models showed that distance measures and angular observables need not be characterized by a unique isotropic relation \cite{Saunders1969Shear}. Modern analyses have confronted Bianchi backgrounds with supernova Hubble diagrams and with the temperature and polarization structure of the cosmic microwave background \cite{Schucker2014BianchiI,Pontzen2007BianchiCMB,Planck2016Isotropy,Saadeh2016Isotropic}. The resulting constraints imply that any coherent late-time anisotropy must be small, but they do not remove the theoretical need to understand how anisotropic propagation enters relativistic observables. On the contrary, controlled models are useful precisely because they identify combinations of observables that respond to the integrated optical geometry rather than merely to a local scale factor.

The appropriate language for this problem is the Sachs optical formalism, in which the expansion, shear, and rotation of a null congruence determine the evolution of an infinitesimal light beam \cite{Sachs:1961zz}. The associated Jacobi map converts an angular separation measured at one event into a physical transverse separation at another, and its matrix structure retains information that is lost when propagation is represented by a single scalar distance. Jacobi fields and optical tidal matrices also provide the rigorous bridge between exact light propagation and the gravitational-lens approximation \cite{Seitz1994LightPropagation}. In Bianchi-I cosmology, the null geodesics and Sachs equations can be solved explicitly, yielding direction-dependent angular-diameter distances, optical shear, and generally non-diagonal Jacobi matrices \cite{Fleury:2014rea,Fleury2016GLCBianchiI}. More broadly, the fact that distinct observables may select different effective optical geometries cautions against identifying all cosmological propagation with one universal distance prescription \cite{Fleury:2013owa}.

This distinction suggests a clean thin-lens decomposition. The compact object produces a localized change in the photon direction, whereas the observer--lens, observer--source, and lens--source propagation is carried by the background Jacobi maps. Such a construction preserves the standard concepts of lens equations, critical curves, and caustics \cite{Schneider:1992bge,Seitz1994LightPropagation}, while allowing the distance factors to become matrices. It also provides a controlled way to compare local theories or compact-object geometries through their weak-field deflection laws, as in parameterized compact-lens formalisms \cite{Keeton2005Formalism,Keeton2006FormalismII}. The underlying geodesic and causal framework remains that of general relativity \cite{Wald:1984rg}; however, no exact global solution describing a compact lens embedded in Kasner spacetime is required. The approximation instead assumes a localized lensing region whose scale and light-crossing time are small compared with those of the cosmological background.

Wormholes offer a particularly informative local lens because their optical response can differ qualitatively from that of ordinary positive-mass compact objects. Traversable Lorentzian wormholes were placed in their modern geometric form by Morris and Thorne, and their causal structure, matter requirements, and energy-condition issues have since been developed extensively \cite{Morris:1988cz,Visser:1995cc,Lobo:2017cay}. The Ellis--Bronnikov solution is among the simplest exact examples: it is a regular, two-ended, static geometry sustained by a phantom scalar field and characterized, in its symmetric massless form, by a throat scale rather than by a Schwarzschild mass monopole \cite{Ellis:1973yv,Bronnikov:1973fh}. In the weak-deflection regime, its leading bending angle is proportional to the inverse square of the physical impact parameter, whereas the Schwarzschild result is proportional to the inverse first power. The two lenses therefore translate the same source--lens--observer geometry into different characteristic angular scales.

The lensing phenomenology of wormholes has been explored from several complementary directions. Early work proposed distinctive lensing events as possible signatures of exotic or negative-mass geometries \cite{Cramer1995Natural}, while subsequent analyses derived lens equations and image properties for a variety of wormhole metrics \cite{Nandi2006WormholeLensing,Dey2008WormholeLensing}. For the Ellis geometry specifically, analytic and exact treatments established the bending-angle structure and its weak-field expansion \cite{Bhattacharya2010Bending,Nakajima2012Deflection}, and more recent calculations have quantified higher-order corrections and their accuracy \cite{Cai2023HigherOrder}. Photometric and astrometric microlensing studies further showed that the Ellis deflection law can produce light curves and centroid trajectories unlike those of conventional point lenses \cite{Abe2010Microlensing,Toki2011Astrometric}. Comparisons of Einstein-ring systems have likewise emphasized that the cubic wormhole scaling and the quadratic Schwarzschild scaling offer, at least in principle, a route to lens discrimination \cite{Tsukamoto:2012xs}.

Strong-field lensing supplies another set of discriminants, although it probes a regime different from the weak-field construction adopted here. General treatments of relativistic images around compact spherical objects and black holes established the role of photon spheres, logarithmic strong-deflection expansions, and higher-order image sequences \cite{Virbhadra2000Schwarzschild,Bozza2002StrongField,Bozza2010BlackHoleLensing}. For Ellis and more general wormholes, strong-deflection analyses have identified the optical role of the throat, photon and antiphoton spheres, additional relativistic images, and differences between symmetric and asymmetric configurations \cite{Tsukamoto2016StrongEllis,Shaikh2019Novel,Shaikh2019Strong,Bronnikov2019Lensing}. These studies demonstrate that the local metric can be filtered into observables in several inequivalent ways. The present work addresses a complementary question: how two different local deflection laws are filtered by one and the same anisotropic cosmological propagation.

The literature thus contains a mature theory of compact-object lensing, a substantial phenomenology of wormhole and black-hole signatures, and an exact optical treatment of homogeneous anisotropic cosmologies. What remains insufficiently isolated is the interplay between these ingredients within a common lens mapping. In particular, calculations performed in asymptotically flat space or with isotropic cosmological distances do not determine whether a directional splitting originates in the local compact object or in the integrated background propagation. Conversely, anisotropic distance studies do not by themselves show how the same Jacobi maps are weighted by deflection laws with different impact-parameter dependence. A comparison in which the background is held fixed while the local lens is changed can separate these effects without attributing the full signal to either sector prematurely.

In this work, we formulate gravitational lensing by localized compact objects in a general diagonal Bianchi-I background using angular Jacobi maps and then specialize the construction to Kasner spacetime. The central null ray is aligned with a principal axis, so the screen map is diagonal and the two transverse directions can be followed analytically. We use the Ellis--Bronnikov wormhole and the Schwarzschild black hole as representative lenses, derive their anisotropic thin-lens equations from the same observer--lens--source propagation, and verify the recovery of the standard isotropic limits. The analysis includes the characteristic axis-aligned angular scales, the complete two-dimensional Jacobian, the critical curves and caustics, and selected Kasner orientations that expose the dependence on the assignment of the directional exponents.

The physical purpose of the comparison is to disentangle the global and local components of the lens map. The directional splitting of the axis-aligned characteristic scales depends on the full sequence of Jacobi maps and, after normalization, is independent of the overall wormhole throat or black-hole mass scale. The exact critical curves provide a distinct diagnostic: their intercepts involve crossed directional responses and can remain nearly circular even when the one-dimensional characteristic scales are appreciably separated. Moreover, the inverse-square Ellis--Bronnikov law and the inverse-linear Schwarzschild law weight the same optical background differently. The resulting contrast shows why characteristic-scale splitting, critical-curve size, and critical-curve morphology should be treated as complementary observables. Because the calculation uses leading weak-deflection laws and a vacuum Kasner background, the results are intended as a controlled proof of principle rather than a direct late-time cosmological model.

It is worth also emphasizing that, although the $\Lambda$CDM model has been remarkably successful in
describing a broad range of cosmological observations, persistent
parameter-level tensions have motivated continued interest in cosmological
scenarios beyond the standard FLRW framework. These include the longstanding
$H_0$ and $S_8$ tensions and recent DESI analyses suggesting possible
departures from a cosmological constant \cite{DESI:2024mwx,DESI:2024uvr,Cai:2026swf}. In parallel,
renewed large-scale tests of statistical isotropy have reopened the question
of whether departures from the cosmological principle should be considered
at a more fundamental level, with recent claims of anisotropic structure in
DESI data \cite{SylosLabini:2026gdf} having prompted an active and
as-yet-unresolved debate \cite{Sawala:2026eah}. Taken together, these
developments illustrate that the assumption of exact isotropy, while
empirically well supported at early times, continues to be actively tested
rather than settled. 

In this context, anisotropic cosmologies provide a natural theoretical framework for investigating how departures from isotropy may influence cosmological observables while preserving large-scale homogeneity. The Kasner solution is adopted here as an analytically tractable vacuum background that isolates the effects of anisotropic optical propagation in a controlled setting. It is not intended as a realistic description of the cosmological background over the redshift range considered. Extending the present formalism to more realistic Bianchi-I cosmologies, for which the directional Jacobi maps generally require numerical evaluation, is left for future work.

This paper is organized as follows. Section~II introduces the diagonal Bianchi-I geometry, specializes it to the Kasner solution, and derives the directional Jacobi maps for propagation along a principal axis. Section~III presents the Ellis--Bronnikov and Schwarzschild lens models, constructs the anisotropic thin-lens equation, and checks the isotropic limit. Section~IV derives the Jacobian matrix and the conditions defining critical curves and caustics. Section~V analyzes the directional characteristic scales, their redshift dependence, and the deformation and size of the exact critical curves. Section~VI summarizes the main conclusions, limitations, and possible extensions of the framework.

\section{Anisotropic cosmological background }

To investigate how cosmological anisotropy modifies gravitational lensing observables, we consider a homogeneous but anisotropic cosmological background described by a Bianchi-I geometry. This class of spacetimes constitutes the simplest anisotropic generalization of the spatially flat FLRW universe and provides a natural framework for isolating directional effects in angular-diameter distances, lens mappings, and image formation. In what follows, we specialize to the Kasner limit, which admits analytical expressions for the redshift and distance relations required in the lensing analysis.

\subsection{Bianchi-I geometry and the Kasner limit}

The simplest spatially homogeneous but anisotropic cosmology is
described by the diagonal Bianchi-I metric
\begin{equation}
    ds^2
    =
    -dt^2
    +a_x^2(t)\,dx^2
    +a_y^2(t)\,dy^2
    +a_z^2(t)\,dz^2,
    \label{eq:bianchi_metric}
\end{equation}
where the three directional scale factors, denoted by $a_i(t)$ with $i=x,y,z$, may evolve independently. We normalize the spatial coordinates at the observation epoch $t_0$
by imposing
\begin{equation}
    a_x(t_0)=a_y(t_0)=a_z(t_0)=1.
\end{equation}

The vacuum Bianchi-I solution is the Kasner spacetime, characterized
by the power-law scale factors
\begin{equation}
    a_i(t)
    =
    \left(\frac{t}{t_0}\right)^{p_i},
    \qquad
    i=x,y,z,
    \label{eq:kasner_scale}
\end{equation}
whose exponents satisfy
\begin{equation}
    p_x+p_y+p_z=1,
    \qquad
    p_x^2+p_y^2+p_z^2=1.
    \label{eq:kasner_constraints}
\end{equation}

We first adopt the axially symmetric configuration
\begin{equation}
    p_x=-\frac{1}{3},
    \qquad
    p_y=p_z=\frac{2}{3}.
    \label{eq:kasner_choice}
\end{equation}
Up to permutations of the spatial axes, this is the unique
non-trivial Kasner solution with two equal exponents. It describes
contraction along the $x$ direction and expansion along the $y$ and
$z$ directions, while retaining a residual symmetry between the two
positive-exponent axes. Choosing the line of sight along $z$ therefore
leaves two inequivalent transverse directions, $x$ and $y$, providing
a particularly simple benchmark for anisotropic lensing. To assess the robustness of the resulting optical signatures beyond this residual symmetry, we also consider the triaxial configurations $(p_x,p_y,p_z)=(-2/7,3/7,6/7)$ and $(-2/7,6/7,3/7)$. In all cases, the central null ray remains aligned with the principal $z$ direction; the additional configurations change the assignment of the Kasner exponents relative to the propagation and transverse axes rather than introducing an oblique line of sight. The large anisotropies considered here should not be interpreted as realistic descriptions of the late-time Universe. They are adopted solely to isolate and amplify the geometrical effects associated with anisotropic optical propagation.

Consider now the central null ray propagating exactly along the
$z$ axis. The spatial translational symmetries imply conservation of
the covariant photon momentum $k_z$, while the photon frequency
measured by comoving observers, denoted by $\omega(t)$,
scales as
\begin{equation}
    \omega(t)\propto \frac{1}{a_z(t)}.
\end{equation}
The redshift $\zeta$, defined as the cosmological redshift of a photon emitted at time $t$ and observed at $t_0$, is consequently
\begin{equation}
    1+\zeta
    =
    \frac{\omega(t)}{\omega(t_0)}
    =
    \frac{a_z(t_0)}{a_z(t)}
    =
    \frac{1}{a_z(t)},
    \label{eq:redshift_kasner}
\end{equation}
where the normalization at $t_0$ has been used. For the Kasner
power law, this relation gives
\begin{equation}
    \frac{t(\zeta)}{t_0}
    =
    (1+\zeta)^{-1/p_z}.
    \label{eq:time_redshift_kasner}
\end{equation}

Along the same axial null ray, $ds^2=0$ implies
\begin{equation}
    \frac{dz_{\rm coord}}{dt}
    =
    \pm\frac{1}{a_z(t)}.
\end{equation}
Here, $z_{\rm coord}$ denotes the comoving spatial coordinate along the propagation axis and should not be confused with the cosmological redshift $\zeta$. In what follows, the positive sign is selected by orienting the $z$ axis along the photon trajectory.
We therefore define the longitudinal coordinate interval, namely the positive comoving coordinate separation along the $z$ direction, between an
emission event at $t(\zeta)$ and the observer as
\begin{equation}
    \chi_z(\zeta)
    \equiv
    z_{\rm coord}(t_0)-z_{\rm coord}\!\left[t(\zeta)\right]
    =
    \int_{t(\zeta)}^{t_0}
    \frac{dt}{a_z(t)}.
    \label{eq:chi_bianchi}
\end{equation}
Here, $t(\zeta)$ is the emission time associated with the observed cosmological redshift $\zeta$, while $t_0$ is the observation time. Thus, $\chi_z(\zeta)$ measures the coordinate, rather than the physical, separation along the central null trajectory.
For $p_z\neq1$, substituting Eq.~\eqref{eq:kasner_scale} yields
\begin{equation}
    \chi_z(\zeta)
    =
    \frac{t_0}{1-p_z}
    \left[
        1-
        (1+\zeta)^{-(1-p_z)/p_z}
    \right].
    \label{eq:chi_kasner}
\end{equation}

Equation~\eqref{eq:chi_kasner} describes the coordinate separation
along the central null geodesic. In an anisotropic spacetime it must
not, by itself, be identified with an angular-diameter distance.
The latter is determined by the evolution of infinitesimal transverse
deviations around the central ray and is therefore encoded in the
Jacobi map derived in the next subsection.

\subsection{Directional Jacobi maps}

In an anisotropic spacetime, the relation between an infinitesimal
angular separation measured at one event and the corresponding
physical separation at another event is described by a Jacobi map,
rather than by a single scalar angular-diameter distance. We restrict
the analysis to a central null ray propagating exactly along the
principal $z$ direction of the diagonal Bianchi-I geometry
\eqref{eq:bianchi_metric}. In this axial configuration, the screen
directions may be chosen along $x$ and $y$, and the Jacobi map is
diagonal,
\begin{equation}
    \boldsymbol{\mathcal D}_{a\to c}
    =
    \begin{pmatrix}
        \mathcal D_{a\to c}^{(x)} & 0\\
        0 & \mathcal D_{a\to c}^{(y)}
    \end{pmatrix}.
    \label{eq:axial_jacobi_map}
\end{equation}
Here $a$ denotes the event at which the rays intersect and their
initial angular separation is measured, whereas $c$ denotes the event
at which their physical transverse separation is evaluated.

To determine the diagonal elements, consider a neighboring ray with
a small angular deviation $\vartheta_i^{(a)}$ from the central
$z$-directed ray at the event $a$. Conservation of the covariant
spatial photon momenta implies, to first order in the deviation,
\begin{equation}
    \frac{k_i}{k_z}
    =
    \frac{a_i(t_a)}{a_z(t_a)}
    \vartheta_i^{(a)},
    \qquad i=x,y.
    \label{eq:momentum_angle_relation}
\end{equation}
Here, $k_\mu=g_{\mu\nu}k^\nu$ denotes the covariant photon four-momentum, with $k_i$, $i=x,y$, representing its transverse spatial components and $k_z$ its component along the central propagation direction.
The transverse geodesic equation then gives
\begin{equation}
    \frac{dx^i}{dt}
    =
    \frac{a_i(t_a)}{a_z(t_a)}
    \frac{a_z(t)}{a_i^2(t)}
    \vartheta_i^{(a)}.
    \label{eq:transverse_geodesic_variation}
\end{equation}
Multiplying the resulting coordinate displacement by the transverse
scale factor at the final event $c$, one obtains
\begin{equation}
    \mathcal D_{a\to c}^{(i)}
    =
    a_i(t_c)
    \frac{a_i(t_a)}{a_z(t_a)}
    \int_{t_c}^{t_a}
    \frac{a_z(t)}{a_i^2(t)}\,dt,
    \qquad i=x,y.
    \label{eq:directional_jacobi_general}
\end{equation}
A derivation of Eq.~\eqref{eq:directional_jacobi_general}, including
the physical normalization of the initial angle and the final screen
separation, is presented in Appendix~\ref{app:directional_jacobi}. It yields the axial
specialization of the Jacobi map in Bianchi-I spacetime, being exact
within the infinitesimal-beam approximation for a central ray aligned
with a principal axis. Generic lines of sight require the full Sachs
optical system and generally lead to a non-diagonal Jacobi map
\cite{Fleury:2014rea}.

For the Kasner scale factors \eqref{eq:kasner_scale}, let
\begin{equation}
    q_i \equiv 1+p_z-2p_i.
    \label{eq:qi_definition}
\end{equation}
For $q_i\neq0$, Eq.~\eqref{eq:directional_jacobi_general} becomes
\begin{equation}
    \mathcal D_{a\to c}^{(i)}
    =
    a_i(t_c)
    \frac{a_i(t_a)}{a_z(t_a)}
    \frac{t_0}{q_i}
    \left[
        \left(\frac{t_a}{t_0}\right)^{q_i}
        -
        \left(\frac{t_c}{t_0}\right)^{q_i}
    \right].
    \label{eq:directional_jacobi_kasner}
\end{equation}
When $q_i=0$, the factor in square brackets divided by $q_i$ is
replaced by
\begin{equation}
    \ln\left(\frac{t_a}{t_c}\right).
\end{equation}

We now introduce the three Jacobi maps required by the thin-lens
construction. Denoting the observer, lens, and source events by
$O$, $L$, and $S$, respectively, the observer--lens and
observer--source maps are
\begin{align}
    D_L^{(i)}
    \equiv
    \mathcal D_{O\to L}^{(i)}
    &=
    a_i(t_L)
    \int_{t_L}^{t_0}
    \frac{a_z(t)}{a_i^2(t)}\,dt,
    \label{eq:DL_directional_correct}
    \\
    D_S^{(i)}
    \equiv
    \mathcal D_{O\to S}^{(i)}
    &=
    a_i(t_S)
    \int_{t_S}^{t_0}
    \frac{a_z(t)}{a_i^2(t)}\,dt,
    \label{eq:DS_directional_correct}
\end{align}
where the normalization $a_i(t_0)=a_z(t_0)=1$ has been used.

The lens--source map must instead be normalized with respect to the
physical angular deviation measured at the lens plane. It is therefore
\begin{equation}
    D_{LS}^{(i)}
    \equiv
    \mathcal D_{L\to S}^{(i)}
    =
    a_i(t_S)
    \frac{a_i(t_L)}{a_z(t_L)}
    \int_{t_S}^{t_L}
    \frac{a_z(t)}{a_i^2(t)}\,dt.
    \label{eq:DLS_directional_correct}
\end{equation}
The factor $a_i(t_L)/a_z(t_L)$ is essential: it converts the physical
angular deviation measured in the orthonormal frame at the lens into
the conserved transverse-to-longitudinal momentum ratio.

For the Kasner background and $q_i\neq0$, these distances read
\begin{align}
    D_L^{(i)}
    &=
    a_i(t_L)\,
    \frac{t_0}{q_i}
    \left[
        1-
        \left(\frac{t_L}{t_0}\right)^{q_i}
    \right],
    \label{eq:DL_kasner_correct}
    \\
    D_S^{(i)}
    &=
    a_i(t_S)\,
    \frac{t_0}{q_i}
    \left[
        1-
        \left(\frac{t_S}{t_0}\right)^{q_i}
    \right],
    \label{eq:DS_kasner_correct}
    \\
    D_{LS}^{(i)}
    &=
    a_i(t_S)
    \frac{a_i(t_L)}{a_z(t_L)}
    \frac{t_0}{q_i}
    \left[
        \left(\frac{t_L}{t_0}\right)^{q_i}
        -
        \left(\frac{t_S}{t_0}\right)^{q_i}
    \right].
    \label{eq:DLS_kasner_correct}
\end{align}

Using the time--redshift relation
\eqref{eq:time_redshift_kasner}, all three quantities can equivalently
be expressed in terms of $\zeta_L$ and $\zeta_S$. The matrices
$\boldsymbol{\mathcal D}_{O\to L}$,
$\boldsymbol{\mathcal D}_{O\to S}$, and
$\boldsymbol{\mathcal D}_{L\to S}$ provide the geometrical input for
the thin-lens equation in the anisotropic background.

\section{Local lens model and anisotropic lens equation}

Having established the directional Jacobi maps associated with the anisotropic cosmological background, we now introduce the local compact object responsible for the gravitational deflection. Throughout this work, we adopt the standard thin-lens approximation, assuming that the physical size of the lensing region and the associated light-crossing time remain negligible compared with the cosmological curvature and expansion scales.

Within this framework, the lens is treated as a localized weak-field perturbation embedded in the anisotropic background. The optical effects of the cosmological geometry are encoded in the directional Jacobi maps derived in the previous section, while the local lensing physics is described by the corresponding weak-field deflection law. The resulting lens equation couples these two ingredients through the observer--lens, observer--source, and lens--source Jacobi maps.

\subsection{Ellis--Bronnikov wormhole}

As a representative example of an exotic compact lens, we consider the Ellis--Bronnikov wormhole, one of the simplest traversable wormhole geometries known in general relativity \cite{Ellis:1973yv}. Its spacetime is characterized by a throat of radius $r_0$ connecting two asymptotically flat regions without the presence of an event horizon. Unlike Schwarzschild black holes, whose lensing properties are governed by a mass monopole, the Ellis wormhole produces light deflection through its nontrivial topology. This leads to a qualitatively different weak-field behavior that has been extensively discussed in the gravitational-lensing literature.

In the weak-field regime, the leading-order deflection angle of a light ray with physical impact parameter $b$ is
\begin{equation}
    \hat\alpha_{\rm WH}(b)
    =
    \frac{\pi r_0^2}{4b^2}
    +
    \mathcal{O}\!\left(\frac{r_0^4}{b^4}\right),
    \label{eq:ellis_deflection}
\end{equation}
where $r_0$ denotes the wormhole throat radius.

Owing to the spherical symmetry of the Ellis--Bronnikov geometry, the deflection in the local lens plane is directed along the physical impact-parameter vector. Thus, the scalar deflection angle in Eq.~\eqref{eq:ellis_deflection} can be promoted to a two-dimensional radial vector as
\begin{equation}
    \hat{\boldsymbol{\alpha}}_{\rm WH}
    =
    \frac{\pi r_0^2}{4b^2}
    \frac{\boldsymbol{\xi}}{b},
    \qquad
    b^2=\xi_x^2+\xi_y^2,
    \label{eq:ellis_vector_deflection}
\end{equation}
where $\boldsymbol{\xi}$ denotes the physical position vector in the lens plane, with components $\boldsymbol{\xi}=(\xi_x,\xi_y)$, while $b=|\boldsymbol{\xi}|$ is its magnitude and therefore the physical impact parameter.

For comparison, the weak-field Schwarzschild deflection is
\begin{equation}
    \hat\alpha_{\rm BH}(b)
    =
    \frac{4M}{b},
    \label{eq:sch_deflection}
\end{equation}
where $M$ denotes the black-hole mass. The distinct impact-parameter dependences,
$\hat\alpha_{\rm WH}\propto b^{-2}$ and
$\hat\alpha_{\rm BH}\propto b^{-1}$,
constitute the key difference between the two lens models. As will be shown below, this distinction leads to different directional Einstein scales and image morphologies when the lenses are embedded in the same anisotropic cosmological background.

\subsection{Lens equation in an anisotropic background}

The directional Jacobi maps derived in the previous section provide the geometrical relation between physical and angular separations in the anisotropic background. Within the thin-lens approximation, the gravitational deflection is assumed to occur instantaneously at the lens plane, while the propagation between observer, lens, and source is entirely encoded in the corresponding Jacobi maps.

Let $\boldsymbol{\theta}=(\theta_x,\theta_y)$ denote the observed angular image position and $\boldsymbol{\beta}=(\beta_x,\beta_y)$ the angular source position. The physical source coordinates are related to $\boldsymbol{\beta}$ through
\begin{equation}
    \boldsymbol{\eta}
    =
    \boldsymbol{\mathcal D}_{O\to S}
    \boldsymbol{\beta},
    \label{eq:beta_definition}
\end{equation}
where $\boldsymbol{\mathcal D}_{O\to S}$ is the Jacobi map connecting the observer and source planes, and $\boldsymbol{\eta}=(\eta_x,\eta_y)$ denotes the physical transverse displacement of the source from the central optical axis, evaluated in the Sachs screen at the source plane.

The anisotropic lens equation can then be written in the compact form
\begin{equation}
    \boldsymbol{\beta}
    =
    \boldsymbol{\theta}
    -
    \boldsymbol{\mathcal D}_{O\to S}^{-1}
    \boldsymbol{\mathcal D}_{L\to S}
    \hat{\boldsymbol{\alpha}}(\boldsymbol{\xi}),
    \label{eq:lens_equation_matrix}
\end{equation}
with the physical lens-plane coordinates given by
\begin{equation}
    \boldsymbol{\xi}
    =
    \boldsymbol{\mathcal D}_{O\to L}
    \boldsymbol{\theta}.
    \label{eq:xi_matrix}
\end{equation}
The Jacobi matrices used here are angular maps. Their relation to
affine-normalized Jacobi matrices and the associated absorption of the
lens redshift factor are discussed in
Appendix~\ref{app:directional_jacobi}.

For the axis-aligned configuration considered here, the Jacobi maps are diagonal,
\[
\boldsymbol{\mathcal D}_{O\to L}
=
\mathrm{diag}
\!\left(
D_L^{(x)},
D_L^{(y)}
\right),
\]
with analogous expressions for
$\boldsymbol{\mathcal D}_{O\to S}$ and
$\boldsymbol{\mathcal D}_{L\to S}$.
Equation~\eqref{eq:lens_equation_matrix} therefore reduces to
\begin{equation}
    \beta_i
    =
    \theta_i
    -
    \frac{D_{LS}^{(i)}}{D_S^{(i)}}
    \hat{\alpha}_i(\boldsymbol{\xi}),
    \qquad i=x,y,
    \label{eq:anisotropic_lens_general}
\end{equation}
which formally resembles the standard isotropic lens equation, except that the distance factors become direction dependent.

The physical coordinates in the lens plane are
\begin{equation}
    \xi_i
    =
    D_L^{(i)}\theta_i,
    \qquad i=x,y,
    \label{eq:xi_directional}
\end{equation}
so that the physical impact parameter becomes
\begin{equation}
    b^2
    =
    \left(D_L^{(x)}\theta_x\right)^2
    +
    \left(D_L^{(y)}\theta_y\right)^2.
    \label{eq:anisotropic_impact}
\end{equation}
Substituting the Ellis--Bronnikov weak-field deflection law,
$\hat{\alpha}_i=(\pi r_0^2/4b^3)\,\xi_i$,
one obtains
\begin{equation}
    \beta_i
    =
    \theta_i
    -
    \frac{D_{LS}^{(i)}}{D_S^{(i)}}
    \frac{\pi r_0^2}{4b^3}
    D_L^{(i)}
    \theta_i,
    \label{eq:anisotropic_ellis_lens}
\end{equation}
with $b$ given by Eq.~\eqref{eq:anisotropic_impact}.

Equations~\eqref{eq:anisotropic_impact} and \eqref{eq:anisotropic_ellis_lens} constitute the fundamental lens mapping employed throughout this work. The anisotropy enters exclusively through the directional Jacobi maps, which modify both the conversion between angular and physical coordinates and the effective lensing efficiency along each principal direction. Consequently, although the local Ellis deflection remains radially symmetric in the lens plane, the resulting lens mapping becomes intrinsically anisotropic.

\subsection{Recovery of the isotropic limit}

Before analyzing the anisotropic lensing signatures, it is important to verify that the formalism correctly reproduces the standard isotropic result. In the isotropic limit, $a_x=a_y=a_z\equiv a(t)$, all directional Jacobi maps become identical,
\[
D_L^{(x)}=D_L^{(y)}\equiv D_L,
\qquad
D_S^{(x)}=D_S^{(y)}\equiv D_S,
\qquad
D_{LS}^{(x)}=D_{LS}^{(y)}\equiv D_{LS}.
\]
Consequently, the Jacobi matrices reduce to scalar multiples of the identity matrix and Eq.~\eqref{eq:lens_equation_matrix} recovers the standard isotropic thin-lens equation.

The physical impact parameter then becomes
\begin{equation}
    b
    =
    D_L\sqrt{\theta_x^2+\theta_y^2},
    \label{eq:isotropic_impact}
\end{equation}
while Eq.~\eqref{eq:anisotropic_ellis_lens} reduces to
\begin{equation}
    \boldsymbol{\beta}
    =
    \boldsymbol{\theta}
    -
    \frac{D_{LS}}{D_S}
    \frac{\pi r_0^2}{4b^3}
    D_L\,\boldsymbol{\theta},
    \label{eq:isotropic_lens}
\end{equation}
which is precisely the weak-field lens equation for the Ellis--Bronnikov wormhole in an isotropic cosmological background.

For perfect source--lens alignment, $\boldsymbol{\beta}=0$, the Einstein ring is recovered. Using $b=D_L\theta_E$, Eq.~\eqref{eq:isotropic_lens} yields
\begin{equation}
    \theta_E^{\rm WH}
    =
    \left(
    \frac{\pi r_0^2}{4}
    \frac{D_{LS}}
         {D_S D_L^2}
    \right)^{1/3},
    \label{eq:isotropic_wh_ring}
\end{equation}
which reproduces the well-known cubic scaling of the Einstein radius for Ellis wormholes.

The recovery of Eq.~\eqref{eq:isotropic_wh_ring} provides a non-trivial consistency check of the anisotropic formalism. It demonstrates that the directional Jacobi-map framework reduces smoothly to the standard isotropic lensing theory when the cosmological anisotropy is removed.

For the Schwarzschild lens, the same isotropic limit gives the standard
Einstein radius
\begin{equation}
    \theta_E^{\rm BH}
    =
    \left(
    4M\frac{D_{LS}}{D_S D_L}
    \right)^{1/2}.
    \label{eq:isotropic_bh_ring}
\end{equation}
Thus, the directional formalism reduces smoothly to the usual
isotropic lens equations for both compact-object models considered in
this work.

\section{Critical curves and caustics}

The anisotropic lens equation derived in the previous section defines
the mapping between the image and source planes. The singularity
structure of this mapping is characterized by its critical curves and
the corresponding caustics. These objects determine the regions of
formally divergent magnification and provide a direct geometric
diagnostic of the combined effects of the local deflection law and the
anisotropic cosmological propagation.

To treat the Ellis--Bronnikov and Schwarzschild lenses within the same
formalism, we write their leading-order weak-field deflection amplitudes as
\begin{equation}
    \hat{\alpha}(b)
    =
    \frac{\mathcal A}{b^n},
    \label{eq:generic_deflection}
\end{equation}
where $\mathcal A$ is the model-dependent deflection coefficient and $n$ determines the power-law dependence on the physical impact parameter $b$. For the Ellis--Bronnikov wormhole, $\mathcal A=\pi r_0^2/4$ and $n=2$, recovering \eqref{eq:ellis_deflection}. For the Schwarzschild black hole, $\mathcal A=4M$ and $n=1$, yielding \eqref{eq:sch_deflection}. Since the deflection vector is radial
in the physical lens plane, its components are
\[
    \hat{\alpha}_i(\boldsymbol{\xi})
    =
    \mathcal A\,
    \frac{\xi_i}{b^{n+1}},
\]
with $b$ given by Eq.~\eqref{eq:anisotropic_impact}.

Defining
\begin{equation}
    Q_i
    \equiv
    \mathcal A
    \frac{D_{LS}^{(i)}}{D_S^{(i)}}
    D_L^{(i)},
    \label{eq:generic_lens_coefficient}
\end{equation}
the component-wise anisotropic lens equation becomes
\begin{equation}
    \beta_i
    =
    \theta_i
    -
    Q_i\theta_i b^{-(n+1)},
    \qquad i=x,y.
    \label{eq:generic_anisotropic_lens}
\end{equation}
Equation~\eqref{eq:generic_anisotropic_lens} reduces to
Eq.~\eqref{eq:anisotropic_ellis_lens} for $n=2$ and to the corresponding
anisotropic Schwarzschild lens equation for $n=1$.

The local properties of the lens mapping are encoded in the Jacobian
matrix
\begin{equation}
    A_{ij}
    =
    \frac{\partial\beta_i}{\partial\theta_j}.
    \label{eq:jacobian_def}
\end{equation}
Using
\[
    b^2
    =
    \sum_{k=x,y}
    \left(D_L^{(k)}\theta_k\right)^2,
\]
one finds
\begin{equation}
    A_{ij}
    =
    \delta_{ij}
    -
    Q_i
    \left[
    \delta_{ij}b^{-(n+1)}
    -
    (n+1)\theta_i
    (D_L^{(j)})^2\theta_j
    b^{-(n+3)}
    \right].
    \label{eq:jacobian_generic_aniso}
\end{equation}
Here, $\delta_{ij}$ denotes the Kronecker delta, with $\delta_{ij}=1$ for $i=j$ and $\delta_{ij}=0$ for $i\neq j$, while $i,j\in\{x,y\}$ label the two transverse screen directions.
The directional Jacobi maps enter both through the coefficients $Q_i$
and through the physical impact parameter. Consequently, even though
the local deflection is radial in the lens plane, the complete lens
mapping is intrinsically anisotropic.

For the Ellis--Bronnikov wormhole, Eq.~\eqref{eq:jacobian_generic_aniso}
reduces to
\begin{equation}
    A_{ij}^{\rm WH}
    =
    \delta_{ij}
    -
    C K_i
    \left[
    \delta_{ij}b^{-3}
    -
    3\theta_i
    (D_L^{(j)})^2\theta_j
    b^{-5}
    \right],
    \label{eq:jacobian_ellis_aniso}
\end{equation}
where
\[
    C=\frac{\pi r_0^2}{4},
    \qquad
    K_i=
    \frac{D_{LS}^{(i)}}{D_S^{(i)}}D_L^{(i)}.
\]
For the Schwarzschild black hole, the analogous expression follows by
setting $n=1$ and $\mathcal A=4M$ in
Eq.~\eqref{eq:jacobian_generic_aniso}.

Critical curves are defined by the singularity condition
\begin{equation}
    \det A=0.
    \label{eq:critical_condition}
\end{equation}
In an isotropic background, this condition recovers the circular
critical curve associated with the Einstein ring. In the anisotropic
case, the directional Jacobi maps modify both its angular scale and its
shape, producing a critical structure aligned with the principal
transverse directions.

The corresponding caustics are obtained by mapping the critical curves
into the source plane through
Eq.~\eqref{eq:generic_anisotropic_lens}. They identify the loci across
which the number of lensed images changes and therefore encode the
source-plane manifestation of the anisotropic optical propagation.
The critical curves for both compact-object models, together with their
axis intercepts and axial ratios, are analyzed in Sec.~V.B.

\section{Expected observables}

The anisotropic lens equation derived above predicts observable departures from the circular Einstein-ring morphology characteristic of isotropic lensing. In particular, the directional dependence of the angular-diameter distances introduces preferred angular scales along the principal Kasner axes, leading to measurable distortions of critical curves and image configurations. We now quantify these effects through a set of simple geometric observables that directly probe the underlying cosmological anisotropy.

\subsection{Directional splitting of characteristic lensing scales}

One of the simplest observables associated with anisotropic lensing is
the splitting of the characteristic angular scale along the two
principal transverse directions. In an isotropic spacetime, a single
Einstein scale characterizes the aligned lensing configuration. In an
anisotropic background, however, the directional dependence of the
optical propagation naturally leads to distinct characteristic scales
along different axes.

For an axis-aligned image configuration, the anisotropic lens equation
admits a characteristic angular scale associated with each principal
direction. For a generic weak-field deflection law of the form
$\hat{\alpha}(b)=\mathcal A\,b^{-n}$, the alignment condition yields
\begin{equation}
\theta_{E,i}
=
\left[
\mathcal A
\frac{D_{LS}^{(i)}}
     {D_S^{(i)}
      \left(D_L^{(i)}\right)^n}
\right]^{\frac{1}{n+1}},
\qquad i=x,y,
\label{eq:generic_axis_scale}
\end{equation}
which generalizes the usual Einstein scale to anisotropic cosmological
backgrounds. These quantities should be interpreted as characteristic
axis-aligned scales obtained from the one-dimensional alignment
condition. In general, they do not coincide with the semiaxes of the
exact critical curve, which is determined instead by the full
two-dimensional condition $\det A=0$.

For the Ellis--Bronnikov wormhole,
$\mathcal A=\pi r_0^2/4$ and $n=2$, giving
\begin{equation}
\theta_{E,i}^{\rm WH}
=
\left[
\frac{\pi r_0^2}{4}
\frac{D_{LS}^{(i)}}
     {D_S^{(i)}
      \left(D_L^{(i)}\right)^2}
\right]^{1/3},
\qquad i=x,y,
\label{eq:axis_einstein_wh}
\end{equation}
whereas for the Schwarzschild black hole,
$\mathcal A=4M$ and $n=1$,
\begin{equation}
\theta_{E,i}^{\rm BH}
=
\left[
4M
\frac{D_{LS}^{(i)}}
     {D_S^{(i)}
      D_L^{(i)}}
\right]^{1/2},
\qquad i=x,y.
\label{eq:axis_einstein_bh}
\end{equation}
Substituting the directional Jacobi maps derived in Sec.~II.B into
Eq. (\eqref{eq:axis_einstein_wh}) yields
\begin{equation}
\theta_{E,i}^{\rm WH}
=
\left[
\frac{\pi r_0^2}{4t_0^2}
\,
\frac{q_i^2}
     {a_i(t_L)a_z(t_L)}
\,
\frac{
L^{q_i}-S^{q_i}
}
{
(1-S^{q_i})
(1-L^{q_i})^2
}
\right]^{1/3},
\label{eq:einstein_scale_kasner}
\end{equation}
where
\[
L\equiv\frac{t_L}{t_0},
\qquad
S\equiv\frac{t_S}{t_0}.
\]
Eq. \eqref{eq:einstein_scale_kasner} illustrates an important
feature of anisotropic lensing. The characteristic angular scales are
not determined solely by the local geometry at the lens position.
Instead, they depend on the complete source--lens--observer optical
propagation through the directional Jacobi maps. Consequently, the
anisotropic response contains both local information about the geometry
at the lens epoch and integrated information about the propagation
history.

To quantify the directional splitting independently of the overall
normalization of the lens, we define
\begin{equation}
\Delta_{\rm aniso}^{X}
=
\frac{
\theta_{E,x}^{X}
-
\theta_{E,y}^{X}
}
{
\frac12
\left(
\theta_{E,x}^{X}
+
\theta_{E,y}^{X}
\right)
},
\qquad
X=\{{\rm WH,BH}\},
\label{eq:anisotropic_splitting}
\end{equation}
which measures the relative separation between the characteristic
scales along the two principal directions. Because the overall lens
normalization cancels identically from
Eq.~\eqref{eq:anisotropic_splitting}, the resulting observable depends
only on the anisotropic optical propagation and on the parameters that
specify the background geometry.


\begin{figure}[htp!]
    \centering
    \includegraphics[width=\linewidth]{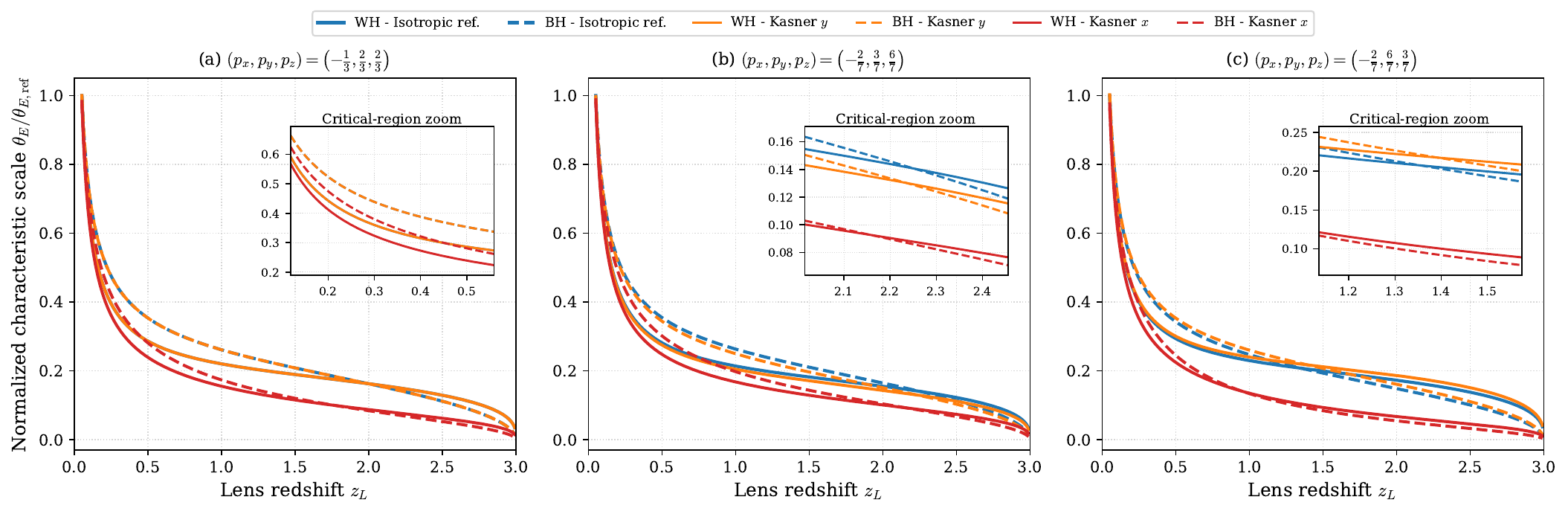}
    \caption{
    Redshift evolution of the normalized characteristic axis-aligned lensing scales for the Ellis--Bronnikov wormhole and Schwarzschild black hole in three Kasner configurations: $(p_x,p_y,p_z)=(-1/3,2/3,2/3)$ in panel (a), $(-2/7,3/7,6/7)$ in panel (b), and $(-2/7,6/7,3/7)$ in panel (c). Solid and dashed curves correspond to the wormhole and black-hole lenses, respectively, while the colors distinguish the matched isotropic reference and the Kasner $x$ and $y$ transverse directions. In each panel, the isotropic reference is defined by the power-law scale factor $a_{\rm iso}(t)=(t/t_0)^{p_z}$, so that it shares the same longitudinal time--redshift relation as the corresponding Kasner background. All curves are normalized by the appropriate isotropic reference value at $\zeta_L=0.05$, with the source fixed at $\zeta_S=3.0$. In panel (a), the equality $p_y=p_z$ causes the Kasner-$y$ curves to coincide with their isotropic counterparts. Panels (b) and (c) show that breaking this residual symmetry and changing the assignment of the Kasner exponents relative to the line of sight modify both directional scales, with the strongest $x$--$y$ separation occurring for the configuration in panel (c).
    }
    \label{fig:thetaE_three_kasner_cases}
\end{figure}

Figure~\ref{fig:thetaE_three_kasner_cases} displays a common decrease of all normalized characteristic scales as $\zeta_L$ approaches the fixed source redshift $\zeta_S=3$. This behavior follows from $D_{LS}^{(i)}\rightarrow0$ when the lens approaches the source, while the differences among the curves arise from the directional combination $D_{LS}^{(i)}/[D_S^{(i)}(D_L^{(i)})^n]$ in Eq.~\eqref{eq:generic_axis_scale}. In all three panels, the central ray remains aligned with the principal $z$ axis; panels~(b) and (c) remove the residual equality between directional exponents but do not correspond to an oblique line of sight. Since each lens model is normalized by its own matched isotropic reference value at $\zeta_L=0.05$, the relative vertical ordering of the solid and dashed curves compares the evolution of the normalized profiles rather than their absolute lensing strengths.

Panel~(a) of Fig.~\ref{fig:thetaE_three_kasner_cases} provides the axially symmetric benchmark. Because $p_y=p_z=2/3$ and the matched isotropic reference uses the same longitudinal exponent, the Kasner-$y$ Jacobi maps reproduce their isotropic counterparts. This explains the exact overlap of the orange and blue curves for each lens model and the apparent absence of separate Kasner-$y$ and reference branches in the main panel and inset. The red curves, associated with the contracting direction $p_x=-1/3$, remain below the overlapping upper branches and become progressively more suppressed as $\zeta_L$ increases. The inset makes the separation between the wormhole and black-hole normalized profiles visible and also shows that their ordering is not preserved throughout the interval.

The triaxial case shown in panel~(b) of Fig.~\ref{fig:thetaE_three_kasner_cases} removes the exact superposition found in panel~(a). With $p_y=3/7<p_z=6/7$, the orange Kasner-$y$ curves lie below the corresponding blue matched-reference curves, whereas the contracting $x$ direction produces the still lower red branches. Thus, for each lens model, the figure exhibits the ordering $\theta_{E,\mathrm{ref}}>\theta_{E,y}>\theta_{E,x}$ over the plotted range. The inset resolves the high-redshift region in which the solid and dashed curves approach and exchange their ordering within the reference, $y$-direction, and $x$-direction families. These crossings reflect the different powers $n=2$ and $n=1$ with which the two deflection laws weight the same directional Jacobi maps.

Panel~(c) of Fig.~\ref{fig:thetaE_three_kasner_cases} uses the same unordered set of Kasner exponents as panel~(b), but assigns the larger positive exponent to the transverse $y$ direction and the smaller one to the line of sight. This permutation changes both the longitudinal relation $t(\zeta)/t_0=(1+\zeta)^{-1/p_z}$ and the integrated transverse propagation. Accordingly, the orange Kasner-$y$ curves now lie above the blue matched-reference curves, while the red Kasner-$x$ branches remain strongly suppressed. The resulting $x$--$y$ separation is visibly the largest of the three configurations. The inset isolates the crossings between the wormhole and black-hole profiles in the reference and $y$-direction families; the $x$ branches exchange their ordering at a lower redshift and are already separated in the magnified interval.

Taken together, the three panels of Fig.~\ref{fig:thetaE_three_kasner_cases} show that the position of a directional branch relative to the matched isotropic reference is controlled by the assignment of the Kasner exponents with respect to the propagation axis, whereas the separation between the solid and dashed profiles reflects the distinct Ellis--Bronnikov and Schwarzschild deflection laws. An intersection between a solid and a dashed curve therefore indicates equality of two separately normalized characteristic scales at that particular redshift; it does not establish an observational degeneracy between the compact objects. Finally, the quantities plotted in Fig.~\ref{fig:thetaE_three_kasner_cases} are axis-aligned scales obtained from the one-dimensional alignment condition. They must not be identified with the semiaxes of the exact critical curves, which follow from the full two-dimensional condition \eqref{eq:critical_condition}.


\subsection{Critical curves and lens discrimination}

The axis-aligned scales introduced in the previous subsection characterize
the solutions of the lens equation under perfect alignment along each
principal direction. A more complete two-dimensional description is
provided by the critical curves, defined by the singularity condition \eqref{eq:critical_condition}. These curves determine
the loci of formally divergent magnification and need not coincide with
the directional alignment scales obtained from $\beta_i=0$.

For the Ellis--Bronnikov lens, it is convenient to define
\[
    G_i
    \equiv
    \frac{\pi r_0^2}{4}
    \frac{D_{LS}^{(i)}}{D_S^{(i)}}
    D_L^{(i)} ,
\]
so that the lens equation reads
$\beta_i=\theta_i-G_i\theta_i b^{-3}$.
Along the horizontal axis, $\theta_y=0$, the tangential singularity is
determined by the transverse Jacobian component $A_{yy}=0$. Conversely,
along the vertical axis, $\theta_x=0$, it is determined by $A_{xx}=0$.
The corresponding critical-curve intercepts are therefore
\begin{equation}
    \theta_{x,{\rm crit}}^{\rm WH}
    =
    \frac{G_y^{1/3}}{D_L^{(x)}},
    \qquad
    \theta_{y,{\rm crit}}^{\rm WH}
    =
    \frac{G_x^{1/3}}{D_L^{(y)}}.
    \label{eq:wh_critical_intercepts}
\end{equation}

This crossed dependence is a distinctive feature of the full
two-dimensional critical condition: the intercept along a given axis is
controlled by the lens response in the transverse direction. It explains
why the directional alignment scales of Sec.~V.A cannot, in general, be
identified with the semiaxes of the critical curve.

For the Schwarzschild lens, defining
\[
    H_i
    \equiv
    4M
    \frac{D_{LS}^{(i)}}{D_S^{(i)}}
    D_L^{(i)} ,
\]
the component-wise lens equation takes the form
$\beta_i=\theta_i-H_i\theta_i b^{-2}$. Its critical-curve intercepts are
\begin{equation}
    \theta_{x,{\rm crit}}^{\rm BH}
    =
    \frac{H_y^{1/2}}{D_L^{(x)}},
    \qquad
    \theta_{y,{\rm crit}}^{\rm BH}
    =
    \frac{H_x^{1/2}}{D_L^{(y)}}.
    \label{eq:bh_critical_intercepts}
\end{equation}

The departures from circular symmetry can be quantified by the axial
ratios
\begin{equation}
    \mathcal R_{\rm WH}
    \equiv
    \frac{\theta_{x,{\rm crit}}^{\rm WH}}
         {\theta_{y,{\rm crit}}^{\rm WH}}
    =
    \frac{D_L^{(y)}}{D_L^{(x)}}
    \left(\frac{G_y}{G_x}\right)^{1/3},
    \qquad
    \mathcal R_{\rm BH}
    \equiv
    \frac{\theta_{x,{\rm crit}}^{\rm BH}}
         {\theta_{y,{\rm crit}}^{\rm BH}}
    =
    \frac{D_L^{(y)}}{D_L^{(x)}}
    \left(\frac{H_y}{H_x}\right)^{1/2}.
    \label{eq:critical_axial_ratios}
\end{equation}
The isotropic limit corresponds to $\mathcal R_{\rm WH}
=\mathcal R_{\rm BH}=1$.

\begin{figure}[htp!]
    \centering
    \includegraphics[width=\linewidth]{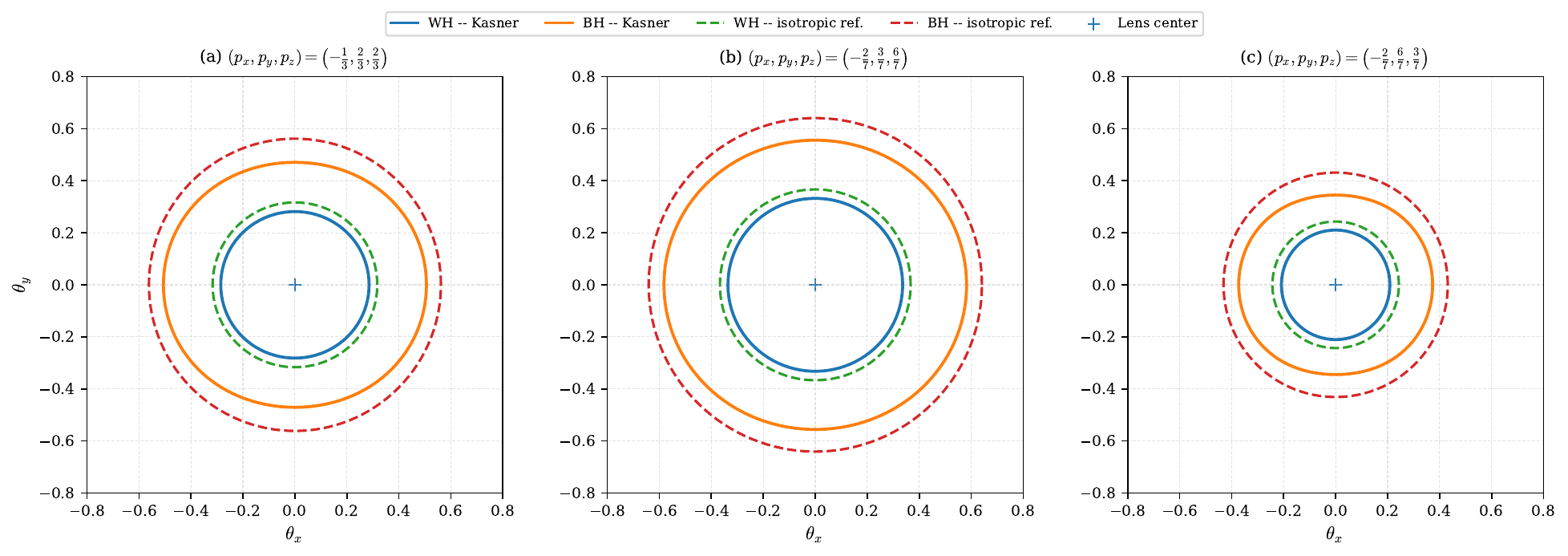}
    \caption{Critical curves for the Ellis--Bronnikov wormhole and the Schwarzschild black hole in three Kasner configurations: $(p_x,p_y,p_z)=(-1/3,2/3,2/3)$ in panel (a), $(-2/7,3/7,6/7)$ in panel (b), and $(-2/7,6/7,3/7)$ in panel (c). The solid blue and orange curves represent the wormhole and black-hole lenses in the corresponding Kasner backgrounds, whereas the dashed green and red circles denote their matched isotropic power-law references, respectively. In each panel, the reference scale factor is $a_{\rm iso}(t)=(t/t_0)^{p_z}$, thereby preserving the longitudinal time--redshift relation of the associated Kasner geometry. The lens and source redshifts are fixed at $\zeta_L\simeq0.59$ and $\zeta_S\simeq3.64$ in all panels, while $2M=r_0=1$. The plus sign marks the lens center.}
    \label{fig:critical_curves_three_kasner_cases}
\end{figure}

Figure~\ref{fig:critical_curves_three_kasner_cases} extends the critical-curve analysis to the three Kasner configurations considered above, allowing the effects of triaxiality and of the assignment of the Kasner exponents relative to the line of sight to be assessed through the full two-dimensional condition \eqref{eq:critical_condition}.

In all three configurations, the Kasner critical curves lie inside their matched isotropic counterparts, showing that the anisotropic propagation reduces the overall angular scale for the adopted lens and source redshifts. This size shift must nevertheless be distinguished from the deformation of the curves. Because the same observed redshifts correspond to different values of $t_L/t_0$ and $t_S/t_0$ when $p_z$ is changed, the variation of the absolute size between panels combines the directional optical response with the modified longitudinal time--redshift relation. The axial ratios provide a cleaner measure of the curve morphology.

Panel~(a) recovers the axially symmetric benchmark. The critical intercepts yield $\mathcal R_{\rm WH}\simeq1.013$ and $\mathcal R_{\rm BH}\simeq1.074$, so both Kasner curves are slightly extended along the $x$ direction. The Ellis--Bronnikov curve remains nearly circular, whereas the Schwarzschild curve exhibits a visibly stronger deformation. This contrast confirms that the weak deformation of the wormhole curve is not simply a consequence of weak background anisotropy; rather, it results from the partial compensation among the directional observer--lens distances and the lensing-efficiency ratios entering the crossed intercepts in Eqs.~\eqref{eq:wh_critical_intercepts} and \eqref{eq:bh_critical_intercepts}.

The triaxial configuration in panel~(b) preserves the orientation of the deformation but changes its magnitude. The corresponding ratios are $\mathcal R_{\rm WH}\simeq1.011$ and $\mathcal R_{\rm BH}\simeq1.047$. Thus, the wormhole curve remains almost circular, while the black-hole elongation along $x$ becomes weaker than in panel~(a). Although this panel has the largest absolute critical curves among the three cases, that increase should not be interpreted as a direct measure of stronger anisotropy, since the matched reference and the cosmic times associated with the fixed redshifts also change when $p_z=6/7$.

Panel~(c) uses the same unordered set of Kasner exponents as panel~(b), but interchanges the positive transverse exponent and the line-of-sight exponent. This permutation produces a qualitatively different response. For the wormhole, $\mathcal R_{\rm WH}\simeq0.991$, corresponding to a small reversal of the preferred elongation from the $x$ direction to the $y$ direction. The Schwarzschild curve instead remains elongated along $x$, with $\mathcal R_{\rm BH}\simeq1.081$, the largest black-hole deformation among the three configurations. The opposite orientation of the two critical curves in this panel provides particularly direct evidence that the same anisotropic Jacobi maps are filtered differently by the $b^{-2}$ and $b^{-1}$ local deflection laws.

The three-panel comparison therefore strengthens the distinction between the axis-aligned characteristic scales and the exact critical morphology. The substantial directional splitting found in Fig.~\ref{fig:thetaE_three_kasner_cases} does not translate into an equally large deformation of the wormhole critical curve, which remains within approximately one percent of circularity in every configuration examined. The black-hole curve is more sensitive to the orientation of the Kasner axes, but its response is not monotonic under permutations of the exponents. Critical-curve size, axial deformation, and directional characteristic-scale splitting must consequently be treated as complementary rather than interchangeable diagnostics of anisotropic optical propagation.

Finally, the normalization $2M=r_0=1$ is adopted to facilitate the comparison of the plotted morphologies. Since the deflection laws employed here are leading weak-field expressions, a quantitative astrophysical application must additionally verify $b/r_0\gg1$ for the wormhole and $b/M\gg1$ for the black hole, or equivalently choose sufficiently small lens scales relative to the cosmological distance scale. The present curves should therefore be interpreted primarily as a controlled comparison within the adopted thin-lens weak-deflection model.


\section{Conclusions}

In this work, we investigated gravitational lensing by localized
compact objects embedded in an anisotropic Bianchi-I cosmological
background. The Ellis--Bronnikov wormhole and the Schwarzschild black
hole were treated within the same thin-lens framework, while the Kasner
solution was adopted as an analytically tractable model of anisotropic
cosmological propagation.

The central element of the construction is the replacement of the
single angular-diameter distance of isotropic cosmology by a
direction-dependent Jacobi map. For a central null ray aligned with a
principal axis of the Bianchi-I geometry, the Jacobi map becomes
diagonal and can be obtained analytically. Its directional components
encode the complete optical propagation between observer, lens, and
source, including the conversion between locally measured angular
deviations and physical transverse separations. The resulting lens
equation therefore couples the local weak-field deflection law of the
compact object to the integrated anisotropic propagation through the
cosmological background.

Specializing to the Kasner geometry, we derived closed-form expressions
for the observer--lens, observer--source, and lens--source Jacobi maps.
These expressions show that the characteristic axis-aligned lensing
scales are not determined solely by the directional scale factors at
the lens epoch. Instead, they depend on the full optical history through
the directional Jacobi maps. This conclusion applies to both compact
objects considered here, although their responses differ because the
Ellis--Bronnikov and Schwarzschild deflection laws scale respectively as
$\hat{\alpha}_{\rm WH}\propto b^{-2}$ and
$\hat{\alpha}_{\rm BH}\propto b^{-1}$.

The redshift evolution displayed in Fig.~\ref{fig:thetaE_three_kasner_cases}
illustrates this distinction across the axially symmetric benchmark and
two triaxial Kasner configurations. The relative positions of the
$x$- and $y$-direction branches depend on the assignment of the Kasner
exponents with respect to the principal propagation axis, while the
wormhole and black-hole profiles respond differently because their
weak-field deflection laws involve distinct powers of the physical
impact parameter. The normalized splitting parameters are independent
of the overall wormhole throat and black-hole mass scales and therefore
isolate the directional response induced by the cosmological propagation.

A complementary observable is provided by the exact critical curves,
which follow from the full two-dimensional condition $\det A=0$.
These curves cannot, in general, be inferred directly from the
axis-aligned characteristic scales because the intercept along one axis
is controlled by the lens response in the transverse direction.
Figure~\ref{fig:critical_curves_three_kasner_cases} shows that the
Ellis--Bronnikov curve remains close to circular in all three
configurations, with $\mathcal R_{\rm WH}\simeq1.013$, $1.011$, and
$0.991$, whereas the Schwarzschild curve is more sensitive to the
assignment of the Kasner exponents, with
$\mathcal R_{\rm BH}\simeq1.074$, $1.047$, and $1.081$. The
contrast confirms that directional splitting and critical-curve
morphology probe different aspects of the anisotropic lens mapping.

The comparisons with the matched isotropic power-law references also
show that anisotropic propagation affects both the shape and the overall
angular size of the critical curves. For the parameters considered here,
the Kasner critical curves lie inside their corresponding isotropic
references in all three configurations. The directional splitting of
the characteristic scales, the axial deformation of the critical
curves, and the shift in their overall size therefore provide
complementary diagnostics of the interplay between large-scale
anisotropic propagation and local compact-object geometry.

The Kasner spacetime should be regarded as a controlled theoretical
laboratory rather than as a realistic description of the late-time
Universe. Its analytical simplicity nevertheless makes it possible to
isolate the optical consequences of anisotropic expansion without the
additional complications introduced by matter sources, evolving shear,
or dark-energy domination. More importantly, although the Kasner spacetime was adopted here as an analytically
tractable testbed, the formalism developed in this work is not restricted to this particular background. Once the directional Jacobi maps are known, either analytically or numerically, the anisotropic lens equation follows immediately for any diagonal Bianchi-I cosmology. The present framework therefore provides a natural extension of the standard thin-lens formalism to homogeneous anisotropic cosmologies. 

Finally, extensions of the present framework may include realistic Bianchi-I cosmologies with time-dependent anisotropy, non-axial lines of sight requiring a fully non-diagonal Jacobi map,
finite-source and ray-tracing calculations, and other compact or exotic
lens geometries. Such developments may help determine whether
direction-dependent lensing observables can provide useful constraints
on cosmological anisotropy while simultaneously distinguishing the local
geometry of compact lenses.

\section*{Acknowledgements}
\noindent CRM would like to thank the Conselho Nacional de Desenvolvimento Cient\'{i}fico e Tecnol\'{o}gico (CNPq) for partial financial support, through grant 301122/2025-3. M. B. Cruz acknowledges financial support from the Conselho Nacional de Desenvolvimento Cient\'{i}fico e Tecnol\'{o}gico (CNPq), through grant 301812/2026-8. RMPN acknowledges financial support from the Funda\c{c}\~{a}o Cearense de Apoio ao Desenvolvimento Cient\'{i}fico e Tecnol\'{o}gico (FUNCAP), under grant BP6-0241-00123.01.00/25.

\bibliographystyle{apsrev4-1}
\bibliography{Ref_1.bib}

\appendix

\section{Derivation of the directional Jacobi maps}
\label{app:directional_jacobi}

We derive here the directional Jacobi maps used in Sec.~II.B for a
central null ray propagating along the principal $z$ direction of the
diagonal Bianchi-I spacetime,
\begin{equation}
ds^2=-dt^2+\sum_{i=x,y,z}a_i^2(t)(dx^i)^2.
\label{eq:app_bianchi}
\end{equation}
The event $a$, at cosmic time $t_a$, denotes the point at which the
initial physical angular deviation is defined, while the event $c$, at
$t_c<t_a$, is the point at which the physical transverse separation is
evaluated.

The comoving observers have four-velocity $u^\mu=\delta^\mu_t$, and an
orthonormal spatial triad is
\begin{equation}
e_{(i)}^\mu=\frac{1}{a_i(t)}\delta^\mu_i.
\label{eq:app_triad}
\end{equation}
For the axial central ray, the transverse vectors $e_{(x)}^\mu$ and
$e_{(y)}^\mu$ span the Sachs screen. Along this ray they are parallel
transported, since
\begin{equation}
k^\nu\nabla_\nu e_{(i)}^i
=
k^t\frac{d}{dt}\left(\frac{1}{a_i}\right)
+
\Gamma^{i}{}_{ti}k^t\frac{1}{a_i}
=0,
\qquad i=x,y.
\label{eq:app_parallel}
\end{equation}
Hence their projections represent physical transverse separations in
the optical screen.

Because the Bianchi-I metric is independent of the spatial coordinates,
the covariant photon momenta
\begin{equation}
k_i=a_i^2(t)\frac{dx^i}{d\lambda}
\label{eq:app_momenta}
\end{equation}
are conserved along the null geodesic. For the unperturbed axial ray,
$k_x=k_y=0$ and the null condition gives
\begin{equation}
|k^t|=\frac{|k_z|}{a_z(t)}.
\label{eq:app_kt}
\end{equation}
The photon frequency measured by comoving observers is therefore
$\omega(t)=|k_z|/a_z(t)$, reproducing the axial redshift relation
$1+\zeta=a_z(t_0)/a_z(t)$.

Consider now a neighboring ray with a small transverse momentum $k_i$,
where $i=x$ or $y$. The physical angular deviation measured at the
initial event $a$ is
\begin{equation}
\vartheta_i^{(a)}
=
\frac{k_i/a_i(t_a)}
     {k_z/a_z(t_a)},
\end{equation}
so that
\begin{equation}
\frac{k_i}{k_z}
=
\frac{a_i(t_a)}{a_z(t_a)}
\vartheta_i^{(a)}.
\label{eq:app_angle_momentum}
\end{equation}
The transverse contributions to the null condition are quadratic in
$\vartheta_i^{(a)}$. Therefore, to first order in the opening angle,
the neighboring and central rays share the same longitudinal evolution,
and
\begin{align}
\frac{dx^i}{dt}
&=
\frac{k_i/a_i^2(t)}
     {k_z/a_z(t)}
\nonumber\\
&=
\frac{a_i(t_a)}{a_z(t_a)}
\frac{a_z(t)}{a_i^2(t)}
\vartheta_i^{(a)}.
\label{eq:app_transverse}
\end{align}

Integrating between $t_c$ and $t_a$ yields the transverse coordinate
separation
\begin{equation}
\Delta x_c^i
=
\frac{a_i(t_a)}{a_z(t_a)}
\left[
\int_{t_c}^{t_a}
\frac{a_z(t)}{a_i^2(t)}\,dt
\right]
\vartheta_i^{(a)}.
\label{eq:app_coordinate_sep}
\end{equation}
The corresponding physical separation in the Sachs screen at the event
$c$ is
\begin{equation}
\xi_c^{(i)}=a_i(t_c)\Delta x_c^i.
\label{eq:app_physical_sep}
\end{equation}
Defining the angular Jacobi map through
$\xi_c^{(i)}=D_{a\to c}^{(i)}\vartheta_i^{(a)}$, one obtains
\begin{equation}
D_{a\to c}^{(i)}
=
a_i(t_c)
\frac{a_i(t_a)}{a_z(t_a)}
\int_{t_c}^{t_a}
\frac{a_z(t)}{a_i^2(t)}\,dt,
\qquad i=x,y.
\label{eq:app_directional_map}
\end{equation}
The factor $a_i(t_a)/a_z(t_a)$ converts the initial physical angular
deviation into the conserved momentum ratio $k_i/k_z$, while
$a_i(t_c)$ converts the final coordinate separation into a physical
screen separation. No weak-anisotropy expansion is involved; the only
linearization is the infinitesimal-beam approximation.

Applying Eq.~\eqref{eq:app_directional_map} to the observer--lens,
observer--source, and lens--source segments gives
\begin{align}
D_L^{(i)}
&=
a_i(t_L)
\int_{t_L}^{t_0}
\frac{a_z(t)}{a_i^2(t)}\,dt,
\label{eq:app_DL}
\\
D_S^{(i)}
&=
a_i(t_S)
\int_{t_S}^{t_0}
\frac{a_z(t)}{a_i^2(t)}\,dt,
\label{eq:app_DS}
\\
D_{LS}^{(i)}
&=
a_i(t_S)
\frac{a_i(t_L)}{a_z(t_L)}
\int_{t_S}^{t_L}
\frac{a_z(t)}{a_i^2(t)}\,dt,
\label{eq:app_DLS}
\end{align}
where $a_i(t_0)=a_z(t_0)=1$ has been used. The additional factor in
$D_{LS}^{(i)}$ is required because the deflection angle is a physical
angle measured in the local orthonormal frame at the lens.

For the Kasner scale factors
$a_i(t)=(t/t_0)^{p_i}$, define
\begin{equation}
q_i\equiv1+p_z-2p_i.
\end{equation}
For $q_i\neq0$,
\begin{equation}
\int_{t_c}^{t_a}
\frac{a_z(t)}{a_i^2(t)}\,dt
=
\frac{t_0}{q_i}
\left[
\left(\frac{t_a}{t_0}\right)^{q_i}
-
\left(\frac{t_c}{t_0}\right)^{q_i}
\right],
\label{eq:app_kasner_integral}
\end{equation}
whereas the limit $q_i=0$ gives
$t_0\ln(t_a/t_c)$. Therefore,
\begin{align}
D_L^{(i)}
&=
a_i(t_L)\frac{t_0}{q_i}
\left[
1-\left(\frac{t_L}{t_0}\right)^{q_i}
\right],
\label{eq:app_DL_kasner}
\\
D_S^{(i)}
&=
a_i(t_S)\frac{t_0}{q_i}
\left[
1-\left(\frac{t_S}{t_0}\right)^{q_i}
\right],
\label{eq:app_DS_kasner}
\\
D_{LS}^{(i)}
&=
a_i(t_S)
\frac{a_i(t_L)}{a_z(t_L)}
\frac{t_0}{q_i}
\left[
\left(\frac{t_L}{t_0}\right)^{q_i}
-
\left(\frac{t_S}{t_0}\right)^{q_i}
\right].
\label{eq:app_DLS_kasner}
\end{align}
These are the expressions quoted in Sec.~II.B. Using
$t(\zeta)/t_0=(1+\zeta)^{-1/p_z}$, they may be written entirely in terms of
$\zeta_L$, $\zeta_S$, and the Kasner exponents.

The maps used in the main text relate a physical angular deviation
directly to a physical transverse separation. If
$\mathscr D_{a\to c}$ denotes an affine-normalized Jacobi matrix, then
\begin{equation}
D_{a\to c}^{(i)}
=
\omega_a\,\mathscr D_{a\to c}^{(i)}.
\label{eq:app_affine_relation}
\end{equation}
Consequently,
\begin{equation}
\boldsymbol{\mathcal D}_{O\to S}^{-1}
\boldsymbol{\mathcal D}_{L\to S}
=
(1+\zeta_L)
\boldsymbol{\mathscr D}_{O\to S}^{-1}
\boldsymbol{\mathscr D}_{L\to S},
\label{eq:app_redshift_factor}
\end{equation}
so that the lens redshift factor is already absorbed into the angular
Jacobi maps employed in Eq.~\eqref{eq:lens_equation_matrix}.

Finally, in the spatially flat FLRW limit,
$a_x=a_y=a_z\equiv a(t)$, Eq.~\eqref{eq:app_directional_map} reduces to
\begin{equation}
D_{a\to c}
=
a(t_c)
\int_{t_c}^{t_a}\frac{dt}{a(t)},
\label{eq:app_flrw_limit}
\end{equation}
which is the standard angular-diameter distance between the two events.
For $a(t)=1$, the Minkowski result
$D_{a\to c}=t_a-t_c$ is recovered.

\end{document}